\documentclass[12pt]{iopart}

\usepackage{iopams}

\usepackage{graphicx}
\usepackage{theorem}
\usepackage{hyperref}

\theoremstyle{plain}
\newtheorem{Thm}{Theorem}

\newtheorem{Prop}[Thm]{Proposition}

\newtheorem{Def}[Thm]{Definition}

\theorembodyfont{\rmfamily}
\newtheorem{Proof}{Proof}

\begin{document}

\title{Solutions to the ultradiscrete Toda molecule equation expressed as minimum weight flows of planar graphs}
\author{Yoichi Nakata}
\address{Graduate School of Mathematical Sciences, The University of Tokyo, 3-8-1 Komaba, Meguro-ku, 153-8914 Tokyo, Japan}
\ead{ynakata@ms.u-tokyo.ac.jp}
\begin{abstract}
We define a function by means of the minimum weight flow on a planar graph and prove that this function solves the ultradiscrete Toda molecule equation, its B\"acklund transformation and the two dimensional Toda molecule equation. The method we employ in the proof can be considered as fundamental to the integrability of ultradiscrete soliton equations.
\end{abstract}

\section{Introduction}

The Box and Ball System(BBS)\cite{TS} is a cellular automaton with genuine soliton-like behaviour in spite of its simple time evolution rule.

A first picture one can use to describe the dynamics of the BBS is to introduce the dependent variable $B^{t}_{j}$ to represent the state of the box at site $j$ and time $t$, defining $B^{t}_{j}=1$ when there is a ball in the box and $B^{t}_{j}=0$ when not. The time evolution rule is then rewritten as
\begin{equation}\label{UTM:evstdBBS}
	B^{t+1}_j = \min \Big( 1 - B^t_j, \sum_{l=-\infty}^{j-1} (B^{t}_l - B^{t+1}_l) \Big)
\end{equation}
where $B^{t}_{j}$ is required to satisfy the following boundary conditions:
\begin{equation}\label{UTM:bcBBS}
	B^t_j = 0 \quad \mbox{for} \ \ |j| \gg 0.
\end{equation}
By employing the dependent variable transformation:
\begin{equation}
	B^t_j = \frac{1}{2} \big( T^{t+1}_j + T^t_{j+1} - T^{t+1}_{j+1} - T^t_j \big),
\end{equation}
the dynamics (\ref{UTM:evstdBBS}) is transformed into
\begin{equation}\label{UTM:uKdV}
	T^{t+2}_{j+1} + T^{t}_{j} = \max \big( T^{t+2}_{j} + T^{t}_{j+1} - 1, T^{t+1}_{j} + T^{t+1}_{j+1} \big),
\end{equation}
which is called the ultradiscrete KdV equation.

\begin{figure}\label{UTM:figure1}
\bigskip
\begin{center}
\includegraphics[bb=0 0 252 36]{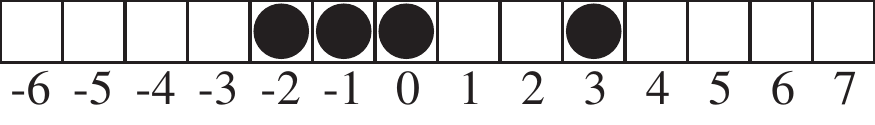}
\caption{An example of a state in the BBS.}
\end{center}
\end{figure}

There is another depiction of the BBS dynamics if one denotes by $E^{t}_{n}$ the length of the $n$-th block of empty boxes and by $Q^{t}_{n}$  the length of the $n$-th block of balls\cite{TNS}. Figure \ref{UTM:figure1} shows an example of the state of the BBS where $E^{t}_{0}=\infty$, $Q^{t}_{1}=3$, $E^{t}_{1}=2$, $Q^{t}_{2}=1$, $E^{t}_{3}=\infty$. The time evolution rule of $Q^{t}_{n}$ and $E^{t}_{n}$ is written as
\begin{eqnarray}
	Q^{t+1}_{n} = \min \Big( \sum_{l=1}^{n} Q^{t}_{l} - \sum_{l=1}^{n-1} Q^{t+1}_{l}, E^{t}_{n-1} \Big) \label{UTM:BBSev2-1} \\
	E^{t+1}_{n} = Q^{t}_{n} + E^{t}_{n} - Q^{t+1}_{n+1}. \label{UTM:BBSev2-2}
\end{eqnarray}
By employing the dependent variable transformations:
\begin{eqnarray}
	Q^{t}_{n} = F^{t}_{n-1} - F^{t}_{n} - F^{t+1}_{n-1} + F^{t+1}_{n} \\
	E^{t}_{n} = F^{t}_{n+1} - F^{t}_{n} - F^{t+1}_{n} + F^{t+1}_{n-1},
\end{eqnarray}
the dynamics (\ref{UTM:BBSev2-1}) and (\ref{UTM:BBSev2-2}) is rewritten as
\begin{equation}
	F^{t}_{n} + F^{t+2}_{n} = \min \big( 2F^{t+1}_{n}, F^{t}_{n+1} + F^{t+2}_{n-1} \big),
\end{equation}
which is called the ultradiscrete Toda molecule equation\cite{NTS1998}.

\bigskip

The dynamics whose time evolution rules can be written in terms of $\min$($\max$) and $\pm$ operators are called ``ultradiscrete systems", which are obtained from discrete soliton equations by the limiting procedure called ``ultradiscretization" \cite{TTMS}. Ultradiscrete soliton equations like the BBS have rich structures such as the existence of N-soliton solutions and an infinite amount of conserved quantities \cite{NTS}, like most ordinary soliton equations. Therefore, ultradiscrete systems are considered to preserve the main characteristics of the soliton equations and it is an interesting problem to obtain the structure of solutions of ultradiscrete soliton equations, as opposed to those of ordinary soliton equations, for example, by means of vertex operators\cite{N}, \cite{N2010} or matrix-based solutions\cite{Nh}, \cite{NT}.

In this paper, we introduce another structure of the soliton solutions of ultradiscrete soliton equations. We first define a function expressed as the minimum weight flow in a planar graph and prove that this function solves the ultradiscrete Toda molecule equation, its B\"acklund transformation and the two dimensional Toda molecule equation. This representation covers all soliton solutions and the properties we employed to prove are considered as the ultradiscretization of the fundamental structure of ordinary soliton equations, such as the Pl\"ucker relation.

Such structured functions given by the weighted flow on the planar graph and their identities are discussed in, for example, \cite{DKK} and \cite{BFZ}. Their applications for discrete or ultradiscrete integrable systems to express the solutions in the form of combinations of the initial values are presented by, for example, \cite{FK1}, \cite{FK2} and \cite{T}.

\section{Minimum weight flow on the skew grid graph}

In this section, we introduce the graph with weighted edges and prove some propositions regarding the properties of this graph.

\begin{Def}
We define the graph $\Gamma$ as follows:
\begin{itemize}
\item It consists of infinitely many vertices labeled $(i, j)$ $i \in \mathbb{Z}, j \in \mathbb{Z}_{\ge 0}$.
\item For $j > 0$, it consists of direct edges:
\begin{itemize}
\item $(i, j)$ to $(i+2, j-1)$, denoted by $d_{i,j}$.
\item $(i, j)$ to $(i, j-1)$, denoted by $v_{i,j}$.
\end{itemize}
\item The weight $w_{edge}$ for an edge $e$ is given by
\begin{equation}
	w_{edge}(e) = \cases{\eta^{i}_{j} &($e=d_{i,j}$)\\0&($e=v_{i,j}$),\\}
\end{equation}
where $\eta^{j}_{j}$ stands for
\begin{equation}
	\eta^{i}_j = i \Omega_j + C_j,
\end{equation}
and the parameters $\Omega_{j}$ satisfy the relations:
\begin{equation}
  \Omega_{1} \le \Omega_{2} \le \ldots \le \Omega_{j} \le \ldots.
\end{equation}
The $C_{j}$ are constants.
\end{itemize}
\end{Def}

\begin{figure}
\begin{center}
\includegraphics[bb=158 423 443 704]{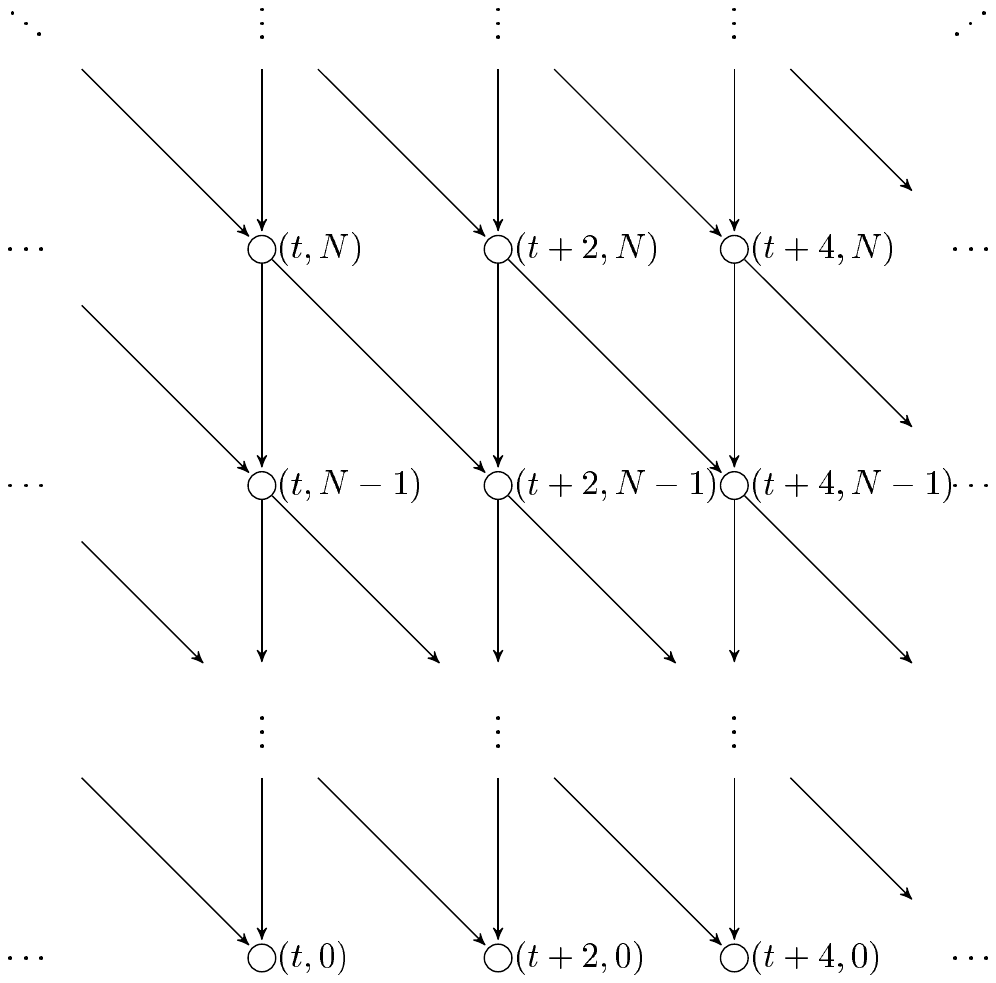}
\caption{A component of $\Gamma$} \label{UTM:figure2}
\end{center}
\end{figure}

By construction, $\Gamma$ is separated into two disjoint planar graphs $\Gamma_{even}$ and $\Gamma_{odd}$ by whether the first component $i$ of the vertex $(i, j)$ is even or odd. Figure \ref{UTM:figure2} shows the planar graph representing $\Gamma_{even}$ or $\Gamma_{odd}$.

\begin{Def}
Let $N>0$ and $0 \le n \le N$. The function $F^{t}_{N, n}$ is defined by
\begin{equation}
  F^{t}_{N, n} =  \cases{\displaystyle \min_{p \in \mathcal{P}^{t}_{N, n}} w(p) &($0 \le n \le N$)\\\infty&(otherwise),\\}
\end{equation}
where $\mathcal{P}^{t}_{N, n}$ is the set of flows on $\Gamma$ with the source $(t, N)$ and the sink $(t+2n, 0)$ and $w(p)$ is the sum of the weights of edges that make up the flow $p$, i.e.
\begin{equation}
	w(p) = \sum_{e \in p} w_{edge}(e).
\end{equation}
We denote $F^{t}_{N,n} = F^{t}_{n}$ for brevity, unless this causes confusion.
\end{Def}
By construction, the set of flows $\mathcal{P}^{t}_{N, n}$ and the edges in the flow $p$ are finite. Since the flow $p$ passes through the edge labeled $d_{t+2(i-1),j}$ only once for each $1 \le i \le n$, we can obtain the sequence of these edges $(d_{t, p_{1}}, \ldots, d_{t-2(n-1), p_{n}})$ or a simpler expression $(p_1, p_2, \ldots, p_n)$ from the flow $p$. Here, by construction, the sequence is decreasing, i.e., $N \ge p_1 > p_2 > \ldots > p_n \ge 1$ and the mapping from the flow $p$ to the decreasing sequence $(p_1, p_2, \ldots, p_n)$ is a bijection. Therefore, we can identify the flow $p$ with the decreasing sequence $(p_1, p_2, \ldots, p_n)$. By virtue of these decreasing sequences, $F^{t}_{n}$ can be rewritten as
\begin{equation}
	F^{t}_{n} = \min_{1\le p_n < p_{n-1} < \ldots < p_1 \le N} \sum_{k=1}^n \eta^{t+2(k-1)}_{p_k}.
\end{equation}

\begin{Prop}\label{UTM:prop1}
For the flows $p \in \mathcal{P}^{t}_{n}$ and $p' \in \mathcal{P}^{t+2}_{n}$, if there exists a number $j$ such that $p_j \ge p'_j + 1$, $p$ and $p'$ share at least one vertex on $\Gamma$.
\end{Prop}

\begin{figure}
\begin{center}
\includegraphics[bb=147 524 332 718]{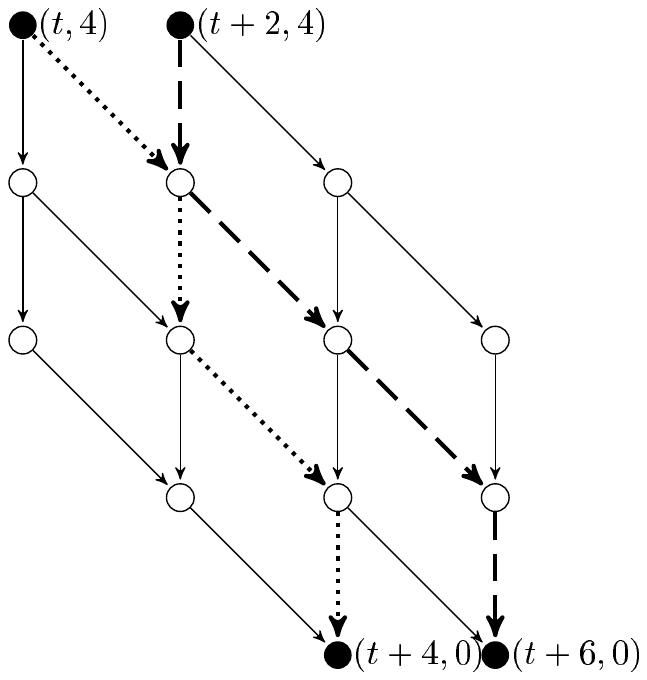}
\caption{An example satisfying Proposition \ref{UTM:prop1}. } \label{UTM:figure4}
\end{center}
\end{figure}

Figure \ref{UTM:figure4} shows an example of this proposition. Two flows $p$ and $p'$ expressed by dotted and dashed line respectively and share the vertex $(t+2, 2)$ for $p_1=3, p'_1=2$.

\begin{Proof}
We consider the first $m$ satisfying the conditions and note that the flow $p$ in $\mathcal{P}^{t}_{n}$ must contain the edge $d_{t+2(m-1), p_m}$, which has the vertex $(t+2m, p_m-1)$. If $m=1$, the flow $p'$ starts at $(t+2, N)$ and moves on the edge $v_{t+2, N}, v_{t+2, N-1}, \ldots, v_{t+2, p_1} \ldots, v_{t+2, p'_1+1}$. Then the edges $d_{t, p_1}$ and $v_{t+2, p_1}$ have the common vertex $(t+2, p_1-1)$. If $m>1$(i.e. $p'_m +1 \le p_m < p_{m-1} \le p'_{m-1}$), by virtue of the same observation, $p'$ contains the edge $v_{t+2m, p_m}$, which has the vertex $(t+2m, p_m-1)$. \hfill $\square$
\end{Proof}

\begin{Prop}\label{UTM:prop2}
Let $p$ and $p'$ be flows which give the minimum weight flow for each source and sink. If $p$ and $p'$ share two vertices $q$ and $q'$, there exist two flows $\tilde{p}$ and $\tilde{p}'$ which go through the same edges from $q$ to $q'$ and keep each weight for each source and sink.
\end{Prop}

\begin{figure}
\begin{center}
\includegraphics[bb=147 524 258 718]{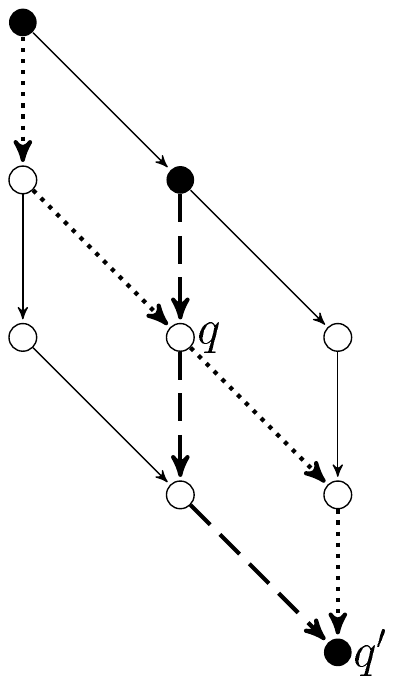}
\caption{An example satisfying Proposition \ref{UTM:prop2}. We can replace the dashed path with the dotted path from $q$ to $q'$ without changing the total weight.} \label{UTM:figure5}
\end{center}
\end{figure}

(See Figure \ref{UTM:figure5} depicting an example for this proposition)

\begin{Proof}
Let $p_{q \to q'}$ the edges where the flow $p$ goes from $q$ to $q'$. If $w(p_{q \to q'}) > w(p'_{q \to q'})$, we replace $p_{q \to q'}$ by $p'_{q \to q'}$ in the flow $p$. The new flow $\tilde{p}$ is that with the same source and sink as $p$ and satisfies $w(\tilde{p}) < w(p)$, which contradicts the definition of $p$. Then, the new path $\tilde{p}$ and $\tilde{p}'=p'$ share the edges from $q$ to $q'$ and has the same weight as $p$ and $p'$ respectively. \hfill $\square$
\end{Proof}

\begin{Prop}\label{UTM:prop3}
Let $p$ and $p'$ be flows which give $F^{t}_{N, m}$ and $F^{t+2l}_{N', n-l}$. If they share at least one vertex, then one has
\begin{equation}\label{UTM:prop3ineq}
	F^{t}_{N, m} + F^{t+2l}_{N', n-l} \ge F^{t}_{N, n} + F^{t+2l}_{N', m-l}.
\end{equation}
\end{Prop}

\begin{figure}
\begin{center}
\includegraphics[bb=147 524 332 718]{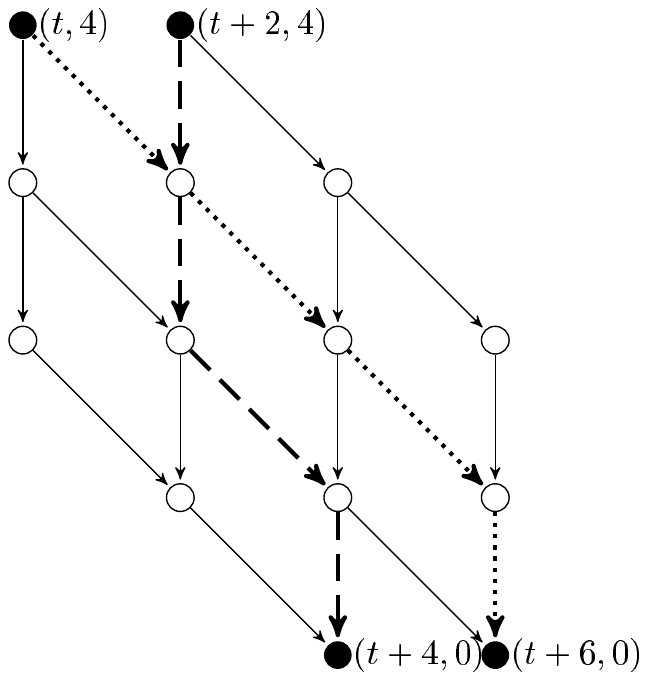}
\includegraphics[bb=147 524 332 718]{figure4.pdf}
\caption{An example illustrating Proposition \ref{UTM:prop3}. By interchanging flows after they intersect, we can obtain new flows without changing the total weight.} \label{UTM:figure6}
\end{center}
\end{figure}

(See Figure \ref{UTM:figure6} depicting an example for this proposition)

\begin{Proof}
By interchanging each orbit after the vertex where the flows intersect (if there are several of such vertices, choose any one of them), we can obtain new paths $\bar{p} \in \mathcal{P}^{t}_{N, n}$ and $\bar{p}' \in \mathcal{P}^{t+2l}_{N', m-l}$. The total weights of the two flows are conserved, before and after the procedure and we obtain (\ref{UTM:prop3ineq}) by virtue of the definition of $F^{t}_{N, n}$.
\end{Proof}

\section{Toda molecule equation}

In this section, we prove the function $F^{t}_{N,n}$ defined in the previous section solves the ultradiscrete Toda molecule equation.

\begin{Thm}\label{UTM:mainThm}
For $0 \le n \le N$, $F^{t}_{n}$ solves the ultradiscrete Toda molecule equation:
\begin{equation}\label{UTM:uTodaM}
	F^{t}_{n} + F^{t+2}_{n} = \min \big( 2F^{t}_{n}, F^{t}_{n+1} + F^{t+2}_{n-1} \big).
\end{equation}
\end{Thm}

When $n=0$ or $n=N$, $F^{t}_{n}$ is linear and $F^{t}_{n+1} + F^{t+2}_{n-1}$ does not contribute to the minimum because  $F^{t}_{-1}=F^{t}_{N+1}=\infty$, so $F$ trivially satisfies the equation (\ref{UTM:uTodaM}). Henceforth, we consider only the case $1 \le n \le N-1$.

\begin{Proof}
We introduce four(two times two) steps to prove.

\begin{description}
\item[(I)] Proof of $F^{t}_{n} + F^{t+2}_{n} \le \min \big( 2F^{t+1}_{n}, F^{t}_{n+1} + F^{t+2}_{n-1} \big)$
\item[(I-i)] Proof of $F^{t}_{n} + F^{t+2}_{n} \le 2 F^{t+1}_{n}$
\end{description}

Let $p \in \mathcal{P}^{t+1}_{n}$ be a flow that gives the minimum weight, i.e. $F^{t+1}_{n} = w(p)$. Here, there exists $p' \in \mathcal{P}^{t}_{n}$ such that $w(p') = \sum_{i=1}^n \eta^{t-2(i-1)}_{p_i} = F^{t+1}_{n} - (\Omega_{p_1} + \Omega_{p_2} + \ldots + \Omega_{p_n})$. Then from the definition of $F^{t}_{n}$ we obtain
\begin{equation}
	F^{t+1}_{n} - (\Omega_{p_1} + \Omega_{p_2} + \ldots + \Omega_{p_n}) = w(p') \ge F^{t}_{n}.
\end{equation}
By virtue of the same discussion, we also obtain
\begin{equation}
	F^{t+1}_{n} + (\Omega_{p_1} + \Omega_{p_2} + \ldots + \Omega_{p_n}) \ge F^{t+2}_{n},
\end{equation}
and adding these inequalities yields the proof.

\begin{description}
\item[(I-ii)] Proof of $F^{t}_{n} + F^{t-2}_{n} \le F^{t}_{n+1} + F^{t-2}_{n-1}$
\end{description}

The flows which give $F^{t}_{n+1}$ and $F^{t+2}_{n-1}$ must intersect because the flow $p$ starts at $(N, t)$ and ends at $(0, t+2n+2)$ and $p'$ starts at $(N, t+2)$ and ends at $(0, t+2n)$. Hence, we obtain the inequality by virtue of Proposition \ref{UTM:prop3}. 

\begin{description}
\item[(II)] Proof of $F^{t}_{n} + F^{t-2}_{n} \ge \min \big( 2F^{t}_{n}, F^{t}_{n+1} + F^{t-2}_{n-1} \big)$
\end{description}

Let $p$ and $p'$ the flows which give $F^{t}_{n}$ and $F^{t+2}_{n}$.

\begin{description}
\item[(II-i)] In case the two flows share at least one vertex.
\end{description}

By virtue of Proposition (\ref{UTM:prop3}), we obtain
\begin{equation}
	F^{t}_{n} + F^{t-2}_{n} \ge F^{t}_{n+1} + F^{t-2}_{n-1}.
\end{equation}

\begin{description}
\item[(II-ii)] In case the two flows do not share any vertices.
\end{description}

By a similar discussion as in (I-i) we obtain
\begin{equation}
	F^{t}_{n} + F^{t+2}_{n} \ge 2 F^{t+1}_{n} - (\Omega_{p_1} + \ldots + \Omega_{p_n}) + (\Omega_{p'_1} + \ldots + \Omega_{p'_n}).
\end{equation}
Here, by the contraposition of Proposition \ref{UTM:prop1}, $p'_j$ must be more than $p_j$ for all $j$(i.e. $\Omega_{p'_j} \ge \Omega_{p_j}$). Then we obtain
\begin{equation}
	F^{t}_{n} + F^{t+2}_{n} \ge 2 F^{t+1}_{n}.
\end{equation}
\hfill $\square$
\end{Proof}

\section{The B\"acklund transformation for the Toda molecule equation}

A B\"acklund transformation is a relation between soliton solutions which have different number of solitons and in general, the procedure to obtain a new solution from a given one by solving some equations.

\begin{Thm}\label{UTM:Backlund}
For $0 \le n \le N$, $F^{t}_{n} = F^{t}_{N,n}$ and $G^{t}_{n} = F^{t}_{N-1,n}$ satisfy the B\"acklund transformation for the Toda molecule equation:
\begin{eqnarray}
	F^{t}_{n} + G^{t}_{n-1} = \min \big( F^{t-1}_{n} + G^{t+1}_{n-1} + \Omega_{N}, F^{t}_{n-1} + G^{t}_{n} \big) \\
	F^{t}_{n} + G^{t+2}_{n} = \min \big( F^{t+1}_{n} + G^{t+1}_{n}, F^{t}_{n+1} + G^{t+2}_{n-1} \big).
\end{eqnarray}
\end{Thm}

\begin{Proof}
As before, we introduce two times two steps for each equation.

\begin{description}
\item[(I)] Proof of $F^{t}_{n} + G^{t}_{n-1} \le \min \big( F^{t-1}_{n} + G^{t+1}_{n-1} + \Omega_{N}, F^{t}_{n-1} + G^{t}_{n} \big)$
\item[(I-i)] Proof of $F^{t}_{n} + G^{t}_{n-1} \le F^{t-1}_{n} + G^{t+1}_{n-1} + \Omega_{N}$
\end{description}

Let $p$ and $p'$ be flows which give $F^{t}_{n}$ and $G^{t}_{n-1}$. We obtain
\begin{equation}
	F^{t-1}_{n} + G^{t+1}_{n-1} \ge F^{t}_{n} + G^{t}_{n} - \Omega_{p_1} + (\Omega_{p'_1} - \Omega_{p_2}) + \ldots + (\Omega_{p'_{n-1}} - \Omega_{p_n}).
\end{equation}
If there exists $m$ such that
\begin{equation}
	\Omega_{p'_{m-1}} < \Omega_{p_m},
\end{equation}
we take $m_0$ to be the smallest number satisfying the above relation, then the two flows must share $(t+2 m_0 - 1, n_0)$. Since they also share the sink $(t+2n, 0)$, by Proposition \ref{UTM:prop2}, we can change the flow $p'$ such that
\begin{equation}\label{UTM:relParam}
	\Omega_{p'_{m-1}} \ge \Omega_{p_m} \ \ \ ( 2 \le m \le n ).
\end{equation}
Hence, we obtain
\begin{equation}
	F^{t-1}_{n} + G^{t+1}_{n-1} \ge F^{t}_{n} + G^{t}_{n-1} - \Omega_{p_1} \ge F^{t}_{n} + G^{t}_{n-1} - \Omega_{N}
\end{equation}
(Figure \ref{UTM:figure8} depicts a visual explanation of the proof).

\begin{figure}
\begin{center}
\includegraphics[bb=147 524 286 718]{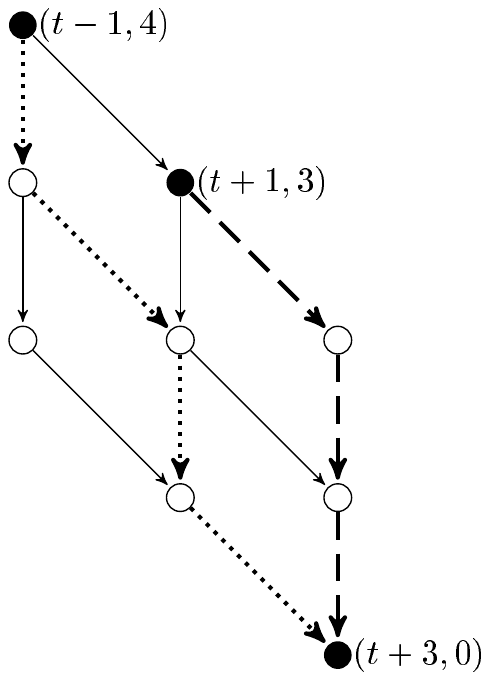}
\includegraphics[bb=147 524 286 718]{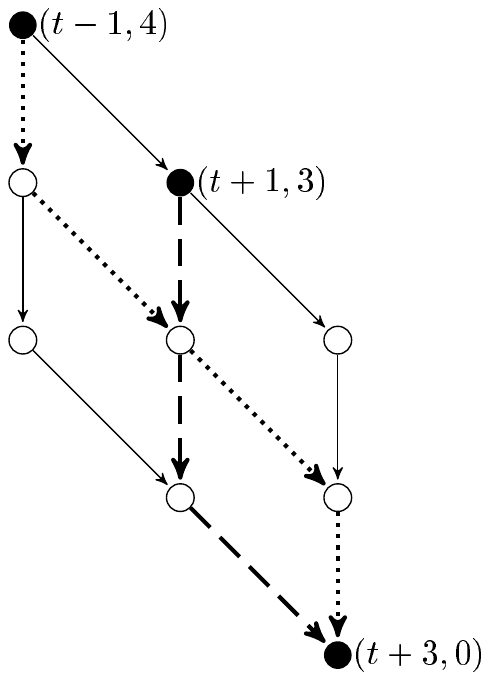}
\includegraphics[bb=147 524 286 718]{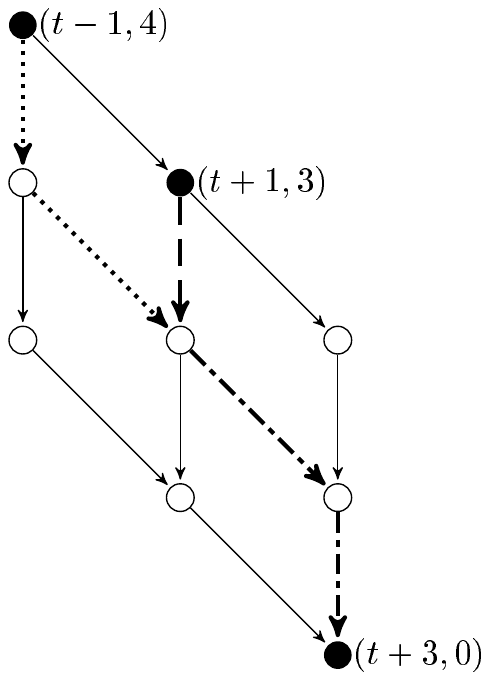}
\caption{An example of the proof of (I-i) in Theorem \ref{UTM:Backlund}. The two flows in the left picture satisfy $\Omega_{p'_1} \ge \Omega_{p_2}$. If the two flows intersect like in the middle, we can change the dotted flow so that is satisfies $\Omega_{p'_1} \ge \Omega_{p_2}$ without changing the total weight as on the right.} \label{UTM:figure8}
\end{center}
\end{figure}

\begin{description}
\item[(I-ii)] Proof of $F^{t}_{n} + G^{t}_{n-1} \le F^{t}_{n-1} + G^{t}_{n}$
\end{description}
We can use Proposition \ref{UTM:prop3} since the flows start at $(t,N)$ and $(t,N-1)$ and terminate at $(t+2n-2, 0)$ and $(t+2n, 0)$ and obtain the inequality.

\begin{description}
\item[(II)] Proof of $F^{t}_{n} + G^{t}_{n-1} \ge \min \big( F^{t-1}_{n} + G^{t+1}_{n-1} + \Omega_{N}, F^{t}_{n-1} + G^{t}_{n} \big)$
\end{description}
Let $p$ and $p'$ the paths which give $F^{t}_{n}$ and $G^{t}_{n-1}$.

\begin{description}
\item[(II-i)] In case the two flows share at least one vertex.
\end{description}
By virtue of Proposition \ref{UTM:prop3}, one then has
\begin{equation}
	F^{t}_{n} + G^{t}_{n-1} \ge F^{t}_{n-1} + G^{t}_{n}.
\end{equation}

\begin{description}
\item[(II-ii)] In case the two flows do not share any vertices.
\end{description}
By the condition, $p$ never has $v_{t, N}$ as its first edge because it cannot go through $(t, N-1)$, which is the source of $p'$. Thus, $p$ must choose $d_{t,N}$, i.e., $p_1=N$, and we can apply the contraposition of Proposition \ref{UTM:prop1} to the flow after passing $d_{t,N}$, i.e. the decreasing sequence $(p_2, p_3, \ldots, p_n) \in \mathcal{P}^{t+2}_{N-1,n-1}$, and $p'$ and obtain
\begin{eqnarray}
	F^{t}_{n} + G^{t}_{n-1} &&\ge F^{t-1}_{n} + G^{t+1}_{n-1} + \Omega_{p_1} + (\Omega_{p_2} - \Omega_{p'_1}) + \ldots + (\Omega_{p_n} - \Omega_{p'_{n-1}}) \nonumber\\
	 &&\ge F^{t-1}_{n} + G^{t+1}_{n-1} + \Omega_{N}.
\end{eqnarray}
(See Figure \ref{UTM:figure9})

\begin{figure}
\begin{center}
\includegraphics[bb=147 524 286 718]{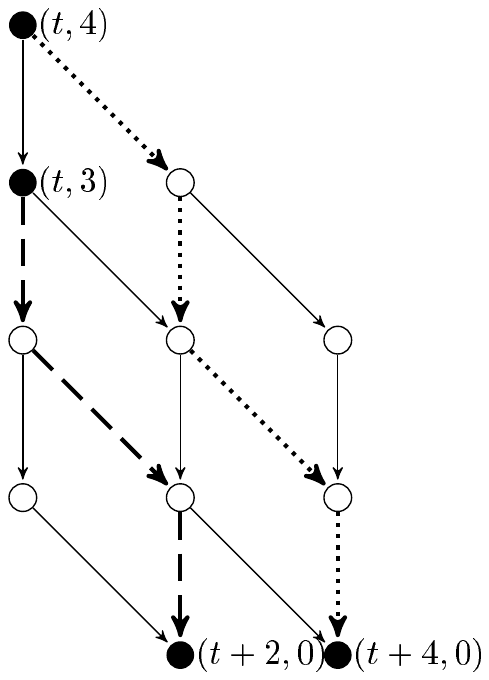}
\caption{An example of the proof of (II-ii) in Theorem \ref{UTM:Backlund}. Due to the condition, $p_1=4$ and after passing $d_{t,4}$ is equivalent to considering the flow in $\mathcal{P}^{t+2}_{3, 1}$.} \label{UTM:figure9}
\end{center}
\end{figure}

\begin{description}
\item[(III)] Proof of $F^{t}_{n} + G^{t+2}_{n} \le \min \big( F^{t+1}_{n} + G^{t+1}_{n}, F^{t}_{n+1} + G^{t+2}_{n-1} \big)$
\item[(III-i)] Proof of $F^{t}_{n} + G^{t+2}_{n} \le F^{t+1}_{n} + G^{t+1}_{n}$
\end{description}
Let $p$ and $p'$ be flows which give $F^{t+1}_{n}$ and $G^{t+1}_{n}$. Since they share the sink $(t+2n, 0)$, by a discussion similar to (I-i) in this proof, we can choose the flows to satisfy
\begin{equation}
	\Omega_{p'_{m}} \ge \Omega_{p_m} \ \ \ ( 1 \le m \le n ).
\end{equation}
Thus, we obtain
\begin{eqnarray}
	F^{t+1}_{n} + G^{t+1}_{n} &&\ge F^{t}_{n} + G^{t+2}_{n} + (\Omega_{p'_1} - \Omega_{p_1}) + \ldots + (\Omega_{p'_n} - \Omega_{p_n}) \nonumber\\
							  &&\ge F^{t}_{n} + G^{t+2}_{n}.
\end{eqnarray}

\begin{description}
\item[(III-ii)] Proof of $F^{t}_{n} + G^{t+2}_{n} \le F^{t}_{n+1} + G^{t+2}_{n-1}$
\end{description}
We can employ Proposition \ref{UTM:prop3} since the flows start at $(t,N)$ and $(t+2,N-1)$ and terminate at $(t+2n+2, 0)$ and $(t+2n, 0)$ and obtain the inequality.

\begin{description}
\item[(IV)] Proof of $F^{t}_{n} + G^{t+2}_{n} \ge \min \big( F^{t+1}_{n} + G^{t+1}_{n}, F^{t}_{n+1} + G^{t+2}_{n-1} \big)$
\end{description}
Let $p$ and $p'$ be the flows which give $F^{t}_{n}$ and $G^{t}_{n-1}$.

\begin{description}
\item[(IV-i)] In the case of two flows sharing at least one vertex.
\end{description}
By virtue of Proposition \ref{UTM:prop3}, then one has
\begin{equation}
	F^{t}_{n} + G^{t+2}_{n} \ge F^{t}_{n+1} + G^{t+2}_{n-1}.
\end{equation}

\begin{description}
\item[(IV-ii)] In the case of two flows not sharing any vertices.
\end{description}
One has:
\begin{equation}
	F^{t+1}_{n} + G^{t+1}_{n} \ge F^{t}_{n} + G^{t+2}_{n-1} + (\Omega_{p_1} - \Omega_{p'_1}) + \ldots + (\Omega_{p_n} - \Omega_{p'_n})
\end{equation}
and by the condition, $p$ never has $d_{t, N}$ as its first edge because it cannot go through $(t+2, N-1)$. Thus, $p$ must choose $v_{t,N}$ and we can employ Proposition \ref{UTM:prop1} for the flow after passing $v_{t,N}$ and $p'$ and obtain
\begin{equation}
	F^{t+1}_{n} + G^{t+1}_{n-1} \ge F^{t}_{n} + G^{t+2}_{n-1}.
\end{equation}
\hfill $\square$
\end{Proof}

\section{Two dimensional Toda molecule equation}

The two dimensional Toda molecule equation is an extension of the Toda molecule equation and the structure of its solution can be thought of as an extension of that of the Toda molecule equation. In this section we first introduce an extended graph and the flow over it and prove that the function given by the minimum weight solves the equation.

\begin{Def}
We define the graph $\tilde{\Gamma}$ as follows:
\begin{itemize}
\item It consists of infinitely many vertices labeled $(i, j, k)$, where $i, j \in \mathbb{Z}, k \in \mathbb{Z}_{\ge 0}$.
\item For $j > 0$, it consists of direct edges:
\begin{itemize}
\item $(i, j, k)$ to $(i+1, j-1, k-1)$, denoted by $d_{i,j,k}$.
\item $(i, j, k)$ to $(i, j, k-1)$, denoted by $v_{i,j,k}$.
\end{itemize}
\item The weight $w_{edge}$ for the edge $e$ is given by
\begin{equation}
	w_{edge}(e) = \cases{\eta^{i,j}_{k} &($e=d_{i,j,k}$)\\0&($e=v_{i,j,k}$).\\}
\end{equation}
Here, $\eta^{l,m}_{j}$ stands for
\begin{equation}
	\eta^{l,m}_k = l P_k - m Q_k + C_k,
\end{equation}
and the parameters $C_{j}$ are constants and $P_k, Q_k$ satisfy the relations:
\begin{eqnarray}
	P_1 \le P_2 \le \ldots \le P_k \le \ldots \\
    (\mbox{or}\ Q_1 \le Q_2 \le \ldots \le Q_k \le \ldots)
\end{eqnarray}
and 
\begin{equation}\label{UTM:cond2dtm}
	\sum_{i=1}^{n} P_{p_{i}} \le \sum_{i=1}^{n} P_{p'_{i}} \Longleftrightarrow \sum_{i=1}^{n} Q_{p_{i}} \le \sum_{i=1}^{n} Q_{p'_{i}}
\end{equation}
for all decreasing sequences $N \ge p_1 > p_2 > \ldots > p_n \ge 1$.
\end{itemize}
\end{Def}

The graph $\tilde{\Gamma}$ is now expressed as an infinite disjoint component consisting of planar graphs. However, we may consider a finite subgraph when treating the flow.

\begin{Def}
Let $N>0$ and $0 \le n \le N$. The function $F^{l,m}_{N, n}$ is defined by
\begin{equation}
  F^{l,m}_{N, n} =  \cases{\displaystyle \min_{p \in \mathcal{P}^{l,m}_{N, n}} w(p) &($0 \le n \le N$)\\\infty&(otherwise),\\}
\end{equation}
where $\mathcal{P}^{l,m}_{N, n}$ is the set of flows on $\tilde{\Gamma}$ with the source $(l, m, N)$ and the sink $(l+n, m-n, 0)$ and $w(p)$ is the sum of the weight of the edges in the flow $p$. We denote $F^{l.m}_{N,n} = F^{l,m}_{n}$ for brevity unless it is confusing.
\end{Def}

By the construction of the flow it should be clear that we can also identify the flow $p$ with the decreasing sequence $( p_1, p_2, \ldots,  p_n )$ and rewritten as
\begin{equation}
	F^{l,m}_{N, n} = \min_{1\le p_n < p_{n-1} < \ldots < p_1 \le N} \sum_{k=1}^n \eta^{l+(k-1), m-(k-1)}_{p_k}.
\end{equation}

\begin{Thm}
The function $F^{l,m}_{n}$ satisfies the ultradiscrete two dimensional Toda molecule equation:
\begin{equation}
	F^{l,m}_{n} + F^{l+1,m-1}_{n} = \min \big( F^{l+1,m}_{n} + F^{l,m-1}_{n}, F^{l,m}_{n+1} + F^{l+1,m-1}_{n-1} \big).
\end{equation}
\end{Thm}

\begin{Proof}
The method is the same as that of Theorem \ref{UTM:mainThm}, and in particular, steps (I-ii) and (II) are completely same. Only the proof of $F^{l,m}_{n} + F^{l+1,m-1}_{n} \le F^{l+1,m}_{n} + F^{l,m-1}_{n}$ requires some additional discussion. 

Let $p$ and $p'$ be the flows which give $F^{l+1,m}_{n}$ and $F^{l,m-1}_{n}$. One has
\begin{equation}
	\!\!\!\!\!\!\!\!\!\!\!\!\!\!\!\! F^{l+1,m}_{n} + F^{l,m-1}_{n} = \sum_{k=1}^n \eta^{l+k-1, m-k+1}_{p_k} + \sum_{k=1}^n P_{p_k} + \sum_{k=1}^n \eta^{l+k-1, m-k+1}_{p'_k} + \sum_{k=1}^n Q_{p'_k}
\end{equation}
If $\sum_{k=1}^n P_{p_k} \ge \sum_{k=1}^n P_{p'_k}$, we obtain
\begin{eqnarray}
	\!\!\!\!\!\!\!\!\!\!\!\!\!\!\!\! F^{l+1,m}_{n} + F^{l,m-1}_{n} &&\ge \sum_{k=1}^n \eta^{l+k-1, m-k+1}_{p_k} + \sum_{k=1}^n \eta^{l+k-1, m-k+1}_{p'_k} + \sum_{k=1}^n P_{p'_k} + \sum_{k=1}^n Q_{p'_k} \\
	&&\ge F^{l,m}_{n} + F^{l-1,m+1}_{n}.
\end{eqnarray}
If $\sum_{k=1}^n P_{p_k} \le \sum_{k=1}^n P_{p'_k}$, by virtue of the condition (\ref{UTM:cond2dtm}), we obtain the same inequality. \hfill $\square$
\end{Proof}

To end this section, let us give some examples satisfying the strong restriction (\ref{UTM:cond2dtm}). Consider the restriction of the parameters $P_{i}$, $Q_{i}$ such that $M P_{i} = Q_{i}$ for $M>0$. Now, the parameters satisfy the restriction (\ref{UTM:cond2dtm}) and the function $F^{l,m}_{n}$ is such that $F^{l+1,m+M}_{n} = F^{l,m}_{n}$ under the restriction, which is the reduction to the dynamics of multi-kind balls and boxes systems \cite{TTM}(In particular, the dynamics reduce to the standard BBS for $M=1$).

\section{Concluding Remarks}

In this paper, we have introduced functions given by minimal weight flows on planar graphs and have proven they yield the ultradiscrete Toda molecule equation and its B\"acklund transformation, for purely structural reasons.

By virtue of the vertex operator for the ultradiscrete soliton equations introduced in \cite{N} and \cite{N2010}, the $N$-soliton solution $F^{t}_{N, n}$ is expressed as
\begin{equation}\label{UTM:utmvertex}
	F^{t}_{N, n} = \min ( F^{t}_{N-1, n}, \eta^{t}_{N} + F^{t+2}_{N, n-1} ),
\end{equation}
which can be interpreted as separating the argument of $\min$ in (\ref{UTM:utmvertex}) according to whether the first edge of the flow is $d_{N,t}$ or $v_{N,t}$. Therefore, the graph representation is more essential for the structure of the solutions of ultradiscrete soliton equations.

\begin{figure}
\begin{center}
\includegraphics[bb=147 572 295 718]{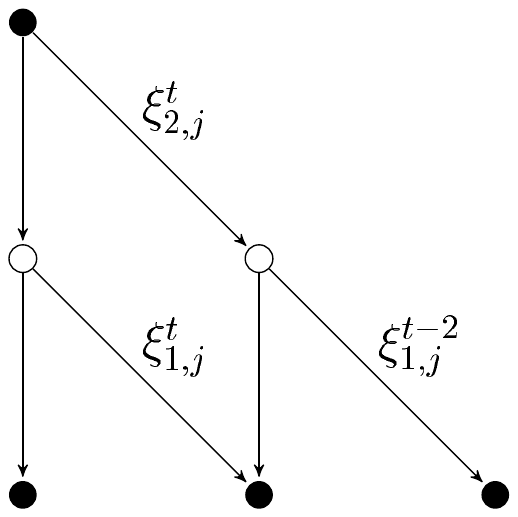}
\caption{An example of the flow representation of a solution of the ultradiscrete KdV equation. The source is the top vertex, the sink is one of the bottom vertices. Numbers over the diagonal edges are weights and $\xi^{t}_{N,j} = t \Omega_{N} - j + C_{N}$.} \label{UTM:figure10}
\end{center}
\end{figure}

The representation is also applicable to that of the ultradiscrete KdV equation (\ref{UTM:uKdV}), which is written as the minimum weight of the flow depicted by Figure \ref{UTM:figure10}. However, it is more difficult to prove that the function given by the minimum weight flow solves the equation relying only on its structure. One of the reasons may be the difference of the relation on which each equation is based, i.e., the discrete KdV equation is based on the Pl\"ucker relation for the Casorati determinant and the discrete Toda molecule equation is based on the Jacobi's formula for a determinant(though it is a special case of the Pl\"ucker relation). It is an interesting problem to clarify why these equations describe the same dynamics in spite of such differences.

It is also an interesting problem to try to discover the direct relationship between the graph structure of the ultradiscrete Toda molecule equation and the determinant solutions of the discrete Toda molecule equation.

\section*{Acknowledge}
The author thanks to Professor T. Tokihiro and Professor R. Willox for helpful comments.

\bibliographystyle{unsrt}
\bibliography{references}

\section*{References}
\begin{thebibliography}{10}

\bibitem{TS}
D.~Takahashi and J.~Satsuma.
\newblock A soliton cellular automaton.
\newblock {\em J. Phys. Soc. Jpn.}, 59:3514--3519, 1990.

\bibitem{TNS}
T.~Tokihiro, A.~Nagai, and J.~Satsuma.
\newblock Proof of solitonical nature of box and ball systems by means of
  inverse ultra-discretization.
\newblock {\em Inverse Problems}, 15(6):1639--1662, 1999.

\bibitem{NTS1998}
A.~Nagai, T.~Tokihiro, and J.~Satsuma.
\newblock {Ultra-discrete Toda molecule equation}.
\newblock {\em Phys. Lett. A}, 244:383--388, 1998.

\bibitem{TTMS}
T.~Tokihiro, D.~Takahashi, J.~Matsukidaira, and J.~Satsuma.
\newblock {From Soliton Equations to Integrable Cellular Automata through a
  Limiting Procedure}.
\newblock {\em Phys. Rev. Lett.}, 76:3247--3250, 1996.

\bibitem{NTS}
A.~Nagai, T.~Tokihiro, and J.~Satsuma.
\newblock Conserved quantities of box and ball system.
\newblock {\em Glasg. Math. J.}, 43A:91--97, 2001.

\bibitem{N}
Y.~Nakata.
\newblock {Vertex operator for the ultradiscrete KdV equation}.
\newblock {\em J. Phys. A: Math. Theor.}, 42:412001 (6pp), 2009.

\bibitem{N2010}
Y.~Nakata.
\newblock {Vertex operator for the non-autonomous ultradiscrete KP equation}.
\newblock {\em J. Phys. A: Math. Theor.}, 43:195201 (8pp), 2010.

\bibitem{Nh}
H.~Nagai.
\newblock {A new expression of soliton solution to the ultradiscrete Toda
  equation}.
\newblock {\em J. Phys. A: Math. Theor.}, 41:235204 (12pp), 2008.

\bibitem{NT}
H.~Nagai and D.~Takahashi.
\newblock {Bilinear equations and Backlund transformation for a generalized
  ultradiscrete soliton solution}.
\newblock {\em J. Phys. A: Math. Theor.}, 43:375202 (13pp), 2010.

\bibitem{DKK}
V.~I. Danilov, A.~V. Karzanov, and G.~A. Koshevoy.
\newblock {\em {Tropical Pl\"ucker functions and their bases}}.
\newblock American Mathematical Society, 2009.

\bibitem{BFZ}
A.~Berenstein, S.~Fomin, and A.~Zelevinsky.
\newblock Parametrizations of canonical bases and totally positive matrices.
\newblock {\em Adv. Math.}, 122:49--149, 1996.

\bibitem{FK1}
P.~D. Francesco and R.~Kedem.
\newblock {Q-systems, heaps, paths and cluster positivity}.
\newblock {\em Comm. Math. Phys.}, 293(3):727--802, 2009.

\bibitem{FK2}
P.~D. Francesco and R.~Kedem.
\newblock {Noncommutative integrability, paths and quasi-determinants}.
\newblock arXiv:1006.4774, 2010.

\bibitem{T}
T. Takagaki
\newblock {Preprint}
\newblock {\em {Koukyuroku, Kyushu University (JAPANESE)}}.

\bibitem{TTM}
T.~Tokihiro, D.~Takahashi, and J.~Matsukidaira.
\newblock {Box and ball system as a realization of ultradiscrete nonautonomous
  KP equation}.
\newblock {\em J. Phys. A: Math. Gen.}, 33:607--619, 2000.

\end{thebibliography}

\end{document}